\documentclass[twocolumn,showpacs,preprintnumbers,amsmath,amssymb,nofootinbib]{revtex4-1}

\usepackage{graphicx}
\usepackage{dcolumn}
\usepackage{bm}
\usepackage{amsfonts}
\usepackage{mathrsfs}
\usepackage{color}
\usepackage{verbatim}

\def\bequ{\begin{equation}}
\def\eequ{\end{equation}}
\def\be{\begin{equation}}
\def\ee{\end{equation}}

\begin{document}


\title{Maxwell perturbations in a cavity with Robin boundary conditions:\\ two branches of modes with spectrum bifurcation on Schwarzschild black holes}
\author{Yunhe Lei}
\author{Mengjie Wang}
\email{Corresponding author: mjwang@hunnu.edu.cn.}
\author{Jiliang Jing}
%
%
%
\affiliation{\vspace{2mm}
Department of Physics and Synergetic Innovation Center for Quantum Effects and Applications, Hunan Normal University,
Changsha, Hunan 410081, P.R. China \vspace{1mm}}
\date{\today}

\begin{abstract}
We perform a systematic study of the Maxwell quasinormal spectrum in a mirror-like cavity following the generic Robin type vanishing energy flux principle, by starting with the Schwarzschild black holes in this paper. It is shown that, for black holes in a cavity, the vanishing energy flux principle leads to \textit{two} different sets of boundary conditions. By solving the Maxwell equations with these two boundary conditions both analytically and numerically, we observe \textit{two} distinct sets of modes. This indicates that the vanishing energy flux principle may be applied not only to asymptotically anti-de Sitter (AdS) black holes but also to black holes in a cavity. In the analytic calculations, the imaginary part of the Maxwell quasinormal modes are derived analytically for both boundary conditions, which match well with the numeric results. While in the numeric calculations, we complete a thorough study on the two sets of the Maxwell spectrum by varying the mirror radius $r_m$, the angular momentum quantum number $\ell$, and the overtone number $N$. In particular, we proclaim that the Maxwell spectrum may \textit{bifurcate} for both modes when the mirror is placed around the black hole event horizon, which is analogous to the spectrum bifurcation effects found for the Maxwell fields on asymptotically AdS black holes. This observation provides another example to exhibit the similarity between black holes in a cavity and the AdS black holes.  
\end{abstract}
\maketitle

\section{Introduction}
Black holes (BHs) are predicted by General Relativity (GR), and the existence of BHs has been confirmed by the observations of gravitational waves~\cite{LIGOScientific:2016aoc,LIGOScientific:2016vlm,LIGOScientific:2016lio,LIGOScientific:2016sjg,LIGOScientific:2017ycc,LIGOScientific:2020iuh} and black hole shadow~\cite{EventHorizonTelescope:2019dse,EventHorizonTelescope:2019ths,EventHorizonTelescope:2019pgp,EventHorizonTelescope:2019ggy}, as the direct evidences. While quasinormal modes (QNMs) play vital roles in both of the aforementioned observations. For the former case, QNMs determine the ringdown phase of gravitational waves~\cite{Ferrari:2007dd,Berti:2018vdi}; and for the latter case, in the eikonal limit, the real part of QNMs is related to the radius of the BH shadow and the imaginary part is related to the amplitude ratio between the $n$th image and the $(n+2)$th image~\cite{Jusufi:2019ltj,Yang:2021zqy}. 

Black hole quasinormal modes have been extensively studied, both in GR and in alternative theories of gravity, for different spin fields on various backgrounds, see for example~\cite{Kokkotas:1999bd,Nollert:1999ji,Berti:2009kk,Konoplya:2011qq} and references therein. To obtain QNMs, one has to solve perturbation equations with physically relevant boundary conditions at the asymptotic regions. At the event horizon, ingoing wave boundary conditions are commonly imposed. Boundary conditions at the other asymptotic region, however, are varied with the asymptotic structure of the background spacetimes. For asymptotically flat (de Sitter) spacetimes, QNMs are defined as outgoing wave boundary conditions at infinity (at the cosmological horizon). For asymptotically anti-de Sitter (AdS) spacetimes, on the other hand, recently we have proposed \textit{the vanishing energy flux principle} to impose boundary conditions. These boundary conditions are Robin type, and lead to novel and interesting results, for example, two distinct sets of QNMs are present for both the Maxwell~\cite{Wang:2015goa,Wang:2015fgp,Wang:2021upj,Wang:2021uix} and the Dirac fields~\cite{Wang:2017fie,Wang:2019qja}, and the Maxwell spectrum may bifurcate~\cite{Wang:2021upj}.

The vanishing energy flux principle originates from the fact that the asymptotic AdS boundary acts like a perfectly reflecting mirror~\cite{Wang:2015goa}. From this viewpoint, one may expect that this principle is also applicable to BHs in a mirror-like cavity, and in this paper we will show that this is \textit{indeed} the case.

Actually the BH-mirror system plays a significant role in understanding various phenomena in BH physics. At the linear level, such system may be employed to construct the BH bomb~\cite{Press:1972zz}, and the analytic treatment indicates that superradiant instabilities may be triggered when the mirror is placed far away from the hole~\cite{Cardoso:2004nk}, which also explains why only small AdS BHs are superradiantly unstable~\cite{Cardoso:2004hs}. At the nonlinear level, the BH-mirror system may be used to explore, for example, the endpoint of superradiant instabilities~\cite{Sanchis-Gual:2015lje} and weakly turbulent instability~\cite{Maliborski:2012gx}. 

Considering the relevance of the BH-mirror system, we shall initialize a systematic study on the quasinormal spectrum of spin fields around BHs in a mirror-like cavity, under \textit{the vanishing energy flux principle}. As the first paper in this direction, here we focus on the Maxwell fields interacting with the Schwarzschild-mirror system. To this end, we first formulate the Maxwell equations on the Schwarzschild background by using the Regge-Wheeler-Zerilli approach~\cite{Regge:1957td, Zerilli:1970se} and then, based on the vanishing energy flux principle, derive \textit{two} sets of explicit boundary conditions at the mirror's location and accordingly obtain \textit{two} branches of quasinormal spectrum. For the pure cavity case, two sets of QNMs turn into two (different) sets of normal modes and which are solved analytically. For the case of BHs in a cavity, the Maxwell QNMs are explored thoroughly by using two different numeric approaches and we present the results by varying the mirror radius $r_m$, the angular momentum quantum number $\ell$  and the overtone number $N$. In particular, we disclose that two distinct sets of modes may \textit{bifurcate}. As a supplement to the numeric calculations, we further compute the Maxwell QNMs \textit{analytically} in the low frequency limit and when the mirror is placed far away from BHs by using an asymptotic matching method.

This paper is organized as follows. In Section~\ref{seceq} we present the Maxwell equations on the Schwarzschild BHs in the Regge-Wheeler-Zerilli formalism, and derive the corresponding boundary conditions explicitly for a BH-mirror system, following the vanishing energy flux principle. In Section~\ref{secnum} we introduce a pseudospectral method and a matrix method, and then apply them in Section~\ref{secres} to explore the Maxwell quasinormal spectrum numerically in detail. Final remarks and conclusions are devoted in Section~\ref{discussion}. Some analytic calculations, including normal modes calculations in the Regge-Wheeler-Zerilli formalism and QNMs calculations in the Teukolsky formalism with low frequency approximations, are left in the Appendixes. 

\section{Maxwell equations and boundary conditions}
\label{seceq}
In this section, we briefly review equations of motion for the Maxwell fields on Schwarzschild BHs in the Regge-Wheeler-Zerilli formalism and present, based on the vanishing energy flux principle, the corresponding boundary conditions at the location of a mirror.

\subsection{The Maxwell equations}
We start by considering the Schwarzschild geometry with the following line element
\begin{equation}
ds^2=f(r)dt^2-\dfrac{1}{f(r)}dr^2-r^2\left(d\theta^2+\sin^2\theta d\varphi^2\right) \;,\label{metric}
\end{equation}
with the metric function
\begin{equation}
f(r)\equiv 1-\dfrac{2M}{r}\;,\label{metricfunc}
\end{equation}
where $M$ is the ADM mass, and the event horizon, determined by $f(r_+)=0$, is given by $r_+=2M$.

The Maxwell equations are given by
\begin{equation}
\nabla_{\nu}F^{\mu\nu}=0\;,\label{Maxwelleq}
\end{equation}
where the field strength tensor is defined as $F_{\mu\nu}=\partial_{\mu}A_{\nu}-\partial_{\nu}A_{\mu}$. In the Regge-Wheeler-Zerilli formalism~\cite{Regge:1957td, Zerilli:1970se}, the vector potential $A_\mu$ may be decomposed in terms of the scalar and vector spherical harmonics~\cite{Ruffini:1973}
\begin{equation}
A_{\mu}=e^{-i\omega t}\sum_{\ell, m}\left(\left[\begin{array}{c} 0 \\
0 \\
a^{\ell m}(r) \boldsymbol {S}_{\ell m}\end{array}\right]+\left[
\begin{array}{c}j^{\ell m}(r)Y_{\ell m}  \\
h^{\ell m}(r)Y_{\ell m} \\
k^{\ell m}(r)\boldsymbol {Y}_{\ell m}  
\end{array}\right]\right)\;,\label{Vpotential}
\end{equation}
where $Y_{\ell m}$ are the scalar spherical harmonics, $\boldsymbol {S}_{\ell m}$ and $\boldsymbol {Y}_{\ell m}$ are the vector spherical harmonics with the definitions
\begin{equation}
\boldsymbol {S}_{\ell m}=
\left(\begin{array}{c} \frac{1}{\sin \theta} \partial_{\varphi}Y_{\ell m}  \\
-\sin \theta \partial_{\theta}Y_{\ell m}\end{array}\right)\;,\;\;\;
\boldsymbol {Y}_{\ell m}=
\left(\begin{array}{c} \partial_{\theta}Y_{\ell m}  \\
\partial_{\varphi}Y_{\ell m}\end{array}\right)\;,\nonumber
\end{equation}
and where $\omega$ is the frequency, $\ell$ is the angular momentum quantum number, $m$ is the azimuthal number. Considering the symmetry of the spherical harmonics under the transformations $(\theta,\varphi)\rightarrow(\pi-\theta,\pi+\varphi)$, the first (second) column in the right hand side of Eq.~\eqref{Vpotential} has parity $(-1)^{\ell+1}$ ($(-1)^\ell$), so that we shall call the former (latter) the axial (polar) modes. 

By substituting Eq.~\eqref{Vpotential} into Eq.~\eqref{Maxwelleq}, one obtains the radial part of the Maxwell equations which, by using the tortoise coordinate 
\begin{equation}
\dfrac{dr_*}{dr}=\dfrac{1}{f(r)}\;,\label{tortoisecoor}
\end{equation}
may be written in the Schr$\ddot{o}$dinger-like form 
\begin{equation}
\left(\frac{d^2}{dr_{*}^2}+\omega^2-\ell(\ell+1)\dfrac{f(r)}{r^2}\right)\Psi(r)=0\;,\label{RWZeq}
\end{equation}
where for axial modes
\begin{equation}
\Psi(r)=a^{\ell m}(r)\;,\nonumber
\end{equation}
and for polar modes
\begin{equation}
\Psi(r)=\dfrac{r^2}{\ell(\ell+1)}\left(-i\omega h^{\ell m}(r)-\dfrac{dj^{\ell m}(r)}{dr}\right)\;.\nonumber
\end{equation}
\subsection{Boundary conditions}
The Maxwell QNMs are determined by Eq.~\eqref{RWZeq} with physically motivated boundary conditions. In this paper, we consider a BH-mirror system, so that one has to impose boundary conditions both at the horizon and at a mirror's location. At the horizon, we impose an ingoing wave boundary condition as usual. By placing a mirror at a finite radius $r_m$ and considering a perfectly reflecting mirror, we require that the energy flux of the Maxwell fields should be vanished at $r_m$, following a generic principle we proposed for asymptotically AdS black holes in~\cite{Wang:2015goa} (see also~\cite{Wang:2016dek,Wang:2016zci}).

To this end, by starting from the energy-momentum tensor of the Maxwell field
\begin{equation}
T_{\mu \nu}=F_{\mu\sigma}F^\sigma_{\;\;\;\nu}+\dfrac{1}{4}g_{\mu\nu}F^2\;,\label{EMTensor}
\end{equation}
where $F^2\equiv F^{\alpha\beta}F_{\alpha\beta}$, one may obtain the spatial part of the radial energy flux, which is given by
\begin{equation}
\mathcal{F}|_r\propto f(r)\Psi(r)\Psi^\prime(r)\;,\label{RWZbc1}
\end{equation}
where $\prime$ denotes a derivative with respect to $r$.
Then the vanishing energy flux principle employed at a finite radius, i.e. $\mathcal{F}|_{r_m}=0$, leads to the following two solutions 
\begin{align}
&\Psi(r)|_{r=r_m}=0\;,\label{RWZbc2-1}\\
&\Psi^\prime(r)|_{r=r_m}=0\;,\label{RWZbc2-2}
\end{align}
where $r_m$ is again the mirror radius in the Schwarzschild coordinates. The above two conditions indicate \textit{two} sets of quasinormal spectrum and, as we will show in the numeric calculations, that they are dissimilar as well. Moreover, the conditions given by Eqs.~\eqref{RWZbc2-1} and~\eqref{RWZbc2-2} are exactly the same with those obtained by treating the mirror as a conductor~\cite{Brito:2015oca}.

\section{Numeric methods}
\label{secnum}
In order to explore the Maxwell spectrum thoroughly in a full parameter space, one has to resort to numeric approaches. In this section, we briefly introduce two numeric methods used in this paper to generate data, $i.e.$ a pseudospectral method~\cite{trefethen2000spectral} and a matrix method~\cite{Lin:2016sch,Lin:2017oag}.

\subsection{Pseudospectral method}
This method has been successfully employed to look for the Maxwell quasinormal spectrum on asymptotically AdS spacetimes~\cite{Wang:2021upj}, and here we follow closely the prescription given in~\cite{Wang:2021upj} but adapted with boundary conditions for the mirror case.

We first, for numeric convenience, make the transformation
\begin{equation}
\Psi=e^{-i\omega r_\ast}\phi\;,\label{spectraltrans}
\end{equation}
where the tortoise coordinate $r_\ast$ is defined in Eq.~\eqref{tortoisecoor}, which brings Eq.~\eqref{RWZeq} from a quadratic eigenvalue problem into a linear eigenvalue problem.

Furthermore, by changing the coordinate from $r$ to $z$ through
\begin{equation}
z=1-2\dfrac{r_m-r}{r_m-r_+}\;,\label{rtoz}
\end{equation}
the integration domain is then transformed from $r\in[r_+,r_m]$ to $z\in[-1,+1]$, and Eq.~\eqref{RWZeq} turns into the form
\begin{equation}
\mathcal{T}_0(z)\phi(z)+\mathcal{T}_1(z,\omega)\phi^\prime(z)+\mathcal{T}_2(z)\phi^{\prime\prime}(z)=0\;,\label{spectraleq1}
\end{equation}
where the functions $\mathcal{T}_j (j=0,1,2)$ may be derived straightforwardly by substituting Eqs.~\eqref{spectraltrans} and~\eqref{rtoz} into Eq.~\eqref{RWZeq}, and where $\prime$ denotes a derivative with respect to $z$. One should note that here $\mathcal{T}_1$ is a linear function of $\omega$, i.e. $\mathcal{T}_1(z,\omega)=\mathcal{T}_{1,0}(z)+\omega\mathcal{T}_{1,1}(z)$.

Then one may discretize Eq.~\eqref{spectraleq1} by introducing the Chebyshev points
\begin{equation}
z_j=\cos\left(\dfrac{j\pi}{n}\right)\;,\;\;\;\;\;\;j=0,1,...,n\;,\label{spectralpoints}
\end{equation}
where $n$ denotes the number of grid points, and obtain an algebraic equation
\begin{equation}
(M_0+\omega M_1)\phi(z)=0\;,\label{spectraleq2}
\end{equation}
where the matrices $M_0$ and $M_1$ are composed of the functions $\mathcal{T}_j (j=0,1,2)$ and the Chebyshev differential matrices~\cite{trefethen2000spectral}.

To solve the eigenvalue equation~\eqref{spectraleq2}, one has to impose physically relevant boundary conditions both at the horizon and at a mirror's location. At the horizon, we impose a regular boundary condition for $\phi$, since an ingoing wave boundary condition for $\Psi$ is guaranteed by Eq.~\eqref{spectraltrans}. At a mirror's position, from Eq.~\eqref{spectraltrans}, boundary conditions given by Eqs.~\eqref{RWZbc2-1} and ~\eqref{RWZbc2-2} are transformed into
\begin{align}
&\phi(1)=0\;,\label{RWZbc3-1}\\
&\phi^\prime(1)-\dfrac{i\omega r_m}{2}\phi(1)=0\;,\label{RWZbc3-2}
\end{align}
where 
\begin{equation}
\phi^\prime(1)\equiv\tfrac{d\phi(z)}{dz}|_{z=1}\;.\nonumber
\end{equation}

\subsection{Matrix method}
The Maxwell quasinormal spectrum may be also calculated by using a matrix method~\cite{Lin:2016sch,Lin:2017oag}, and here we briefly describe this method.

Similarly to the pseudospectral method introduced in the above, we first make the transformation given by Eq.~\eqref{spectraltrans} so that an ingoing wave boundary condition at the horizon is satisfied for $\Psi$ automatically.

Then we take a coordinate transformation
\begin{equation}
x=\frac{r-r_+}{r_m-r_+}\;,\label{rtox}
\end{equation}  
which brings the integration domain from $r\in[r_+,r_m]$ to $x\in[0,1]$. We further implement the transformation from $\phi$ to $\chi$ through
\begin{equation}
\chi(x)=x\phi(x)\;,\label{phitochi}
\end{equation}
such that the boundary condition at the event horizon becomes
\begin{equation}
\chi(0)=0\;,\label{BCchih}
\end{equation}
and then Eq.~\eqref{RWZeq} becomes
\begin{equation}
\mathcal{\tilde{T}}_0(x,\omega)\chi(x)+\mathcal{\tilde{T}}_1(x,\omega)\chi'(x)+\mathcal{\tilde{T}}_2(x)\chi''(x)=0\;,\label{MaxEqMatrix}
\end{equation}
where the functions $\tilde{\mathcal{T}}_j (j=0,1,2)$ may be derived directly by substituting Eqs.~\eqref{spectraltrans},~\eqref{rtox} and~\eqref{phitochi} into Eq.~\eqref{RWZeq}. Note that here $\mathcal{\tilde{T}}_j (j=0,1)$ are linear functions of $\omega$, $i.e.$ $\mathcal{\tilde{T}}_j(x,\omega)=\mathcal{\tilde{T}}_{j,0}(x)+\omega\mathcal{\tilde{T}}_{j,1}(x)$.

To discretize Eq.~\eqref{MaxEqMatrix} by using the matrix method, it is used to introduce equally spaced grid points in the internal $[0,1]$. The corresponding differential matrices may be constructed by using the Taylor series to expand the function $\chi(x)$ around each grid point. Then Eq.~\eqref{MaxEqMatrix} becomes an algebraic equation in the matrix form  
\begin{equation}
(\tilde{M}_0+\omega \tilde{M}_1)\chi(x)=0\;,\label{MaxEqmatrix2}
\end{equation}
where $\tilde{M}_0$ and $\tilde{M}_1$ are matrices composed by the functions $\tilde{T}_j (j=0,1,2)$ and the corresponding differential matrices, and the explicit form of these matrices may be found in~\cite{Lin:2016sch,Lin:2017oag}.  

To solve the eigenfrequencies from Eq.~\eqref{MaxEqmatrix2}, apart from the boundary condition at the horizon given in Eq.~\eqref{BCchih}, one has to impose boundary conditions at a mirror's location. Considering the transformations we made in Eqs.~\eqref{spectraltrans} and~\eqref{phitochi}, boundary conditions given by Eqs.~\eqref{RWZbc2-1} and~\eqref{RWZbc2-2} are then transformed into  
\begin{align}
&\chi(1)=0\;,\label{RWZmatrixbc-1}\\
&\chi'(1)-\left(1+i\omega r_m\right)\chi(1)=0\;,\label{RWZmatrixbc-2}
\end{align}
where
\begin{equation}
\chi^\prime(1)\equiv\tfrac{d\chi(x)}{dx}|_{x=1}\;.\nonumber
\end{equation}

\section{Results}
\label{secres}
With the numeric methods illustrated above at hand, one may obtain the Maxwell quasinormal spectrum on Schwarzschild BHs in a cavity. In the numeric calculations we take the event horizon $r_+=1$ to measure all other physical quantities. Moreover, we use $\omega_1$ ($\omega_2$) to represent QNMs corresponding to the first (second) boundary condition given by Eq.~\eqref{RWZbc2-1} (Eq.~\eqref{RWZbc2-2}), and introduce $\ell$ and $N$ to denote the angular momentum quantum number and the overtone number.

\begin{figure*}
\begin{center}
\begin{tabular}{c}
\hspace{-4mm}\includegraphics[clip=true,width=0.326\textwidth]{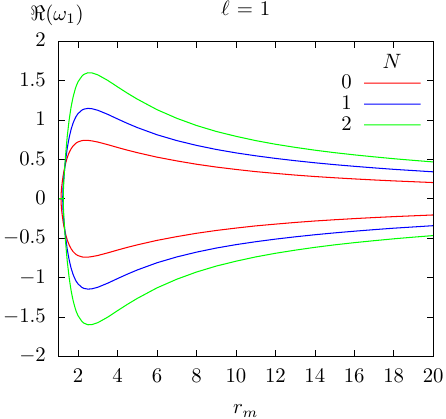}\hspace{12mm}\includegraphics[clip=true,width=0.326\textwidth]{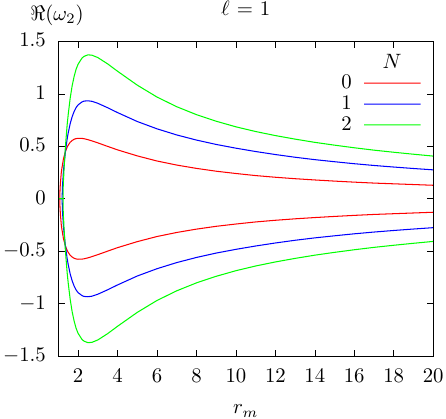}
\\
\vspace{2mm}
\\
\hspace{-2mm}\includegraphics[clip=true,width=0.330\textwidth]{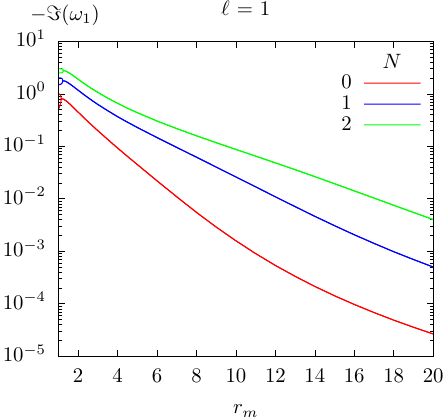}\hspace{11mm}\includegraphics[clip=true,width=0.330\textwidth]{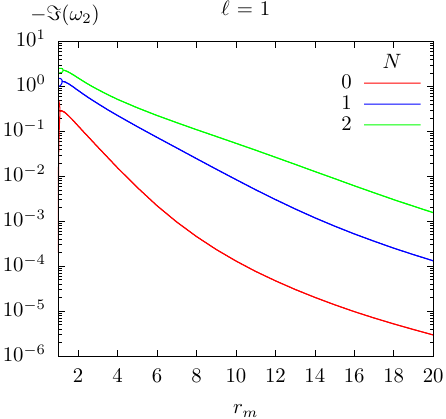}
\end{tabular}
\end{center}
\caption{\label{Fig_ell1variousN} Real (top) and imaginary (bottom) parts of the Maxwell QNMs on Schwarzschild BHs in a cavity vs the mirror radius $r_m$, for $\ell=1$ and $N=0$ (red), $1$ (blue), $2$ (green), with the first (left) and second (right) boundary conditions. Note that in the bottom panels for imaginary part of QNMs with both boundary conditions, we use the semilogarithmic coordinates.}
\end{figure*}

\begin{figure}
\begin{center}
\begin{tabular}{c}
\hspace{-4mm}\includegraphics[clip=true,width=0.386\textwidth]{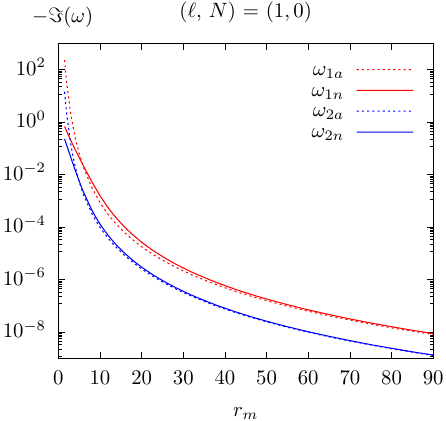}
\end{tabular}
\end{center}
\caption{\label{Fig_comp} A comparison for the imaginary part of the Maxwell QNMs between analytic results with the first ($-\Im\omega_{1a}$) and second ($-\Im\omega_{2a}$) boundary conditions and numeric data with the first ($-\Im\omega_{1n}$) and second ($-\Im\omega_{2n}$) boundary conditions. Here we show the results in terms of mirror radius $r_m$ in the semilogarithmic coordinate, by taking $\ell=1$ and $N=0$ as an example. Notice that the larger mirror radius the better agreement between analytic and numeric results. Also note that, with fixed $r_m$, the agreement with the second boundary is better than the agreement with the first boundary.} 
\end{figure}

We first display both real and imaginary parts of the Maxwell QNMs in terms of the mirror radius $r_m$ with both boundary conditions in Fig.~\ref{Fig_ell1variousN}, by taking $\ell=1$ and $N=0, 1, 2$ as examples. From this figure, we confirm that the vanishing energy flux principle is applicable to a BH-mirror system and leads to \textit{two} different branches of modes. It shows clearly that, for both boundary conditions with fixed $\ell$ and $N$, the magnitude of real (imaginary) part of QNMs first increases up to a maximum and then decreases (always increases), as the mirror approaches the event horizon. By increasing the mirror radius $r_m$, the magnitude of real part of QNMs for both boundary conditions asymptotes to the corresponding normal modes in a pure cavity, given by Eqs.~\eqref{normalmodes1} and~\eqref{normalmodes2} as calculated in Appendix.~\ref{app1}, while the imaginary part for both modes is computed perturbatively by using a standard analytic matching method in Appendix~\ref{secana}. Here we compare analytic results with numeric data in Fig.~\ref{Fig_comp} by varying the mirror radius $r_m$ for $\ell=1$ and $N=0$, and find a good agreement between analytic and numeric results for large $r_m$. Moreover, we notice that, with fixed $r_m$, the agreement with the second boundary is better than the agreement with the first boundary. This comparison not only verifies the validity of the analytic method but also checks our numeric data.

\begin{figure*}
\begin{center}
\begin{tabular}{c}
\hspace{-4mm}\includegraphics[clip=true,width=0.326\textwidth]{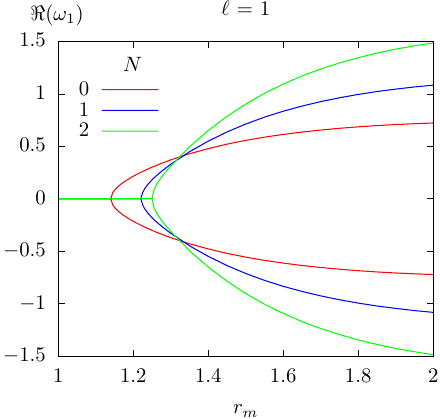}\hspace{12mm}\includegraphics[clip=true,width=0.326\textwidth]{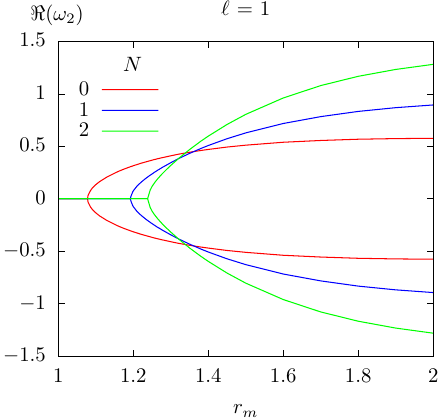}
\\
\vspace{2mm}
\\
\hspace{-2mm}\includegraphics[clip=true,width=0.316\textwidth]{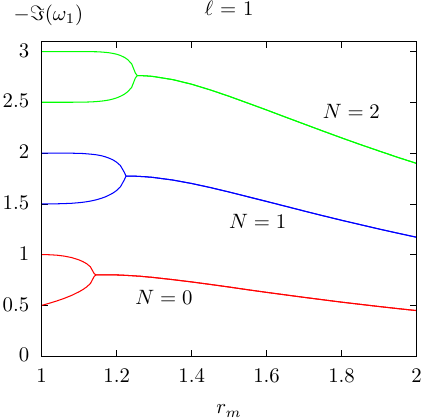}\hspace{14.5mm}\includegraphics[clip=true,width=0.316\textwidth]{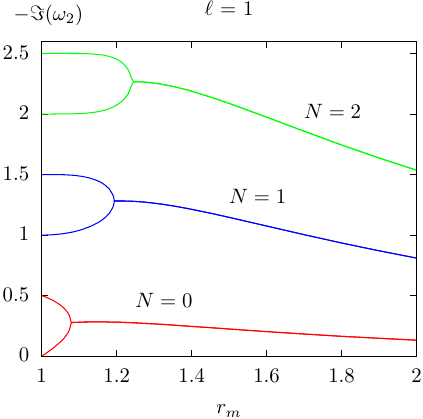}
\end{tabular}
\end{center}
\caption{\label{Fig_ell1variousN_zoom} This figure is obtained by zooming Fig.~\ref{Fig_ell1variousN} around the event horizon to better display the spectrum bifurcation of the Maxwell fields on Schwarzschild BHs in a cavity for both boundary conditions. Note that the critical mirror radius with the first (second) boundary condition are $r_{m1}^c \approx 1.14028$ ($r_{m2}^c \approx 1.07825$) for $N=0$, $r_{m1}^c \approx 1.22161$ ($r_{m2}^c \approx 1.19323$) for $N=1$ and $r_{m1}^c \approx 1.25037$ ($r_{m2}^c \approx 1.23932$) for $N=2$.}
\end{figure*}

By zooming Fig.~\ref{Fig_ell1variousN} around the event horizon, one observes an absorbing feature in Fig.~\ref{Fig_ell1variousN_zoom} that the Maxwell spectrum may \textit{bifurcate} for both modes. To be precise, as shown in Fig.~\ref{Fig_ell1variousN_zoom}, we find out that for both boundary conditions when the real part of QNMs becomes \textit{zero}, which determines a critical mirror radius $r_m^c$,\footnote{Note that, by fixing all other parameters, the critical mirror radius with the first boundary $r_{m1}^c$ is greater than the counterpart with the second boundary $r_{m2}^c$. } the imaginary part \textit{branches off} into to two sets of modes when $r_m$ is smaller than $r_m^c$. This phenomenon, we dubbed as the \textit{mode split} effect, has also been reported for the Maxwell fields on asymptotically AdS BHs~\cite{Wang:2021upj}. By using the same terminology introduced in~\cite{Wang:2021upj}, we term, in magnitude, the larger (smaller) one of the two sets of modes as the upper (lower) mode. The upper (lower) mode, for both boundary conditions, increases (decreases) as $r_m$ approaches the event horizon, and it varies weakly with $r_m$ when the mirror radius is smaller than and away from the critical mirror radius. In particular, when the mirror radius $r_m$ is close to the event horizon, one may achieve a very simple structure of the modes, that is the upper (lower) mode with the first boundary condition asymptotes to $-(N+1)i$ [$-(N+0.5)i$], while with the second boundary the upper (lower) mode asymptotes to $-(N+0.5)i$ [$-Ni$].   


The impact of the overtone number $N$ on the spectrum varies with the mirror position. When the mirror radius is greater than and away from the critical mirror radius, as shown in Fig.~\ref{Fig_ell1variousN}, with a fixed $r_m$ for both boundary conditions, the magnitude for both real and imaginary parts of QNMs increase as $N$ increases; while when the mirror is placed around the critical mirror radius, as shown in Fig.~\ref{Fig_ell1variousN_zoom}, with a fixed $r_m$ for both boundary conditions, the magnitude for real part of QNMs decreases but the imaginary part increases (for both upper and lower modes) as $N$ increases. Moreover, as shown in Fig.~\ref{Fig_ell1variousN_zoom}, the critical mirror radius for both boundary conditions increases as $N$ increases.

\begin{figure*}
\begin{center}
\begin{tabular}{c}
\hspace{-4mm}\includegraphics[clip=true,width=0.326\textwidth]{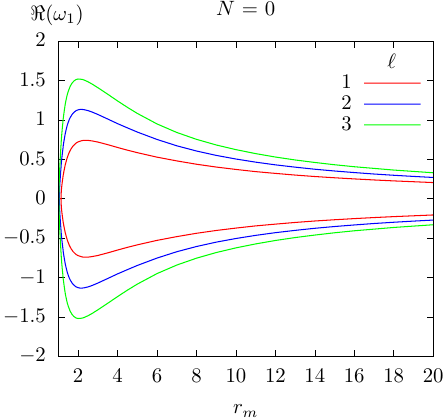}\hspace{12mm}\includegraphics[clip=true,width=0.326\textwidth]{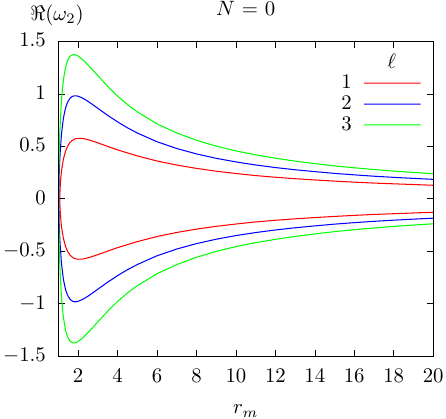}
\\
\vspace{2mm}
\\
\hspace{-2mm}\includegraphics[clip=true,width=0.326\textwidth]{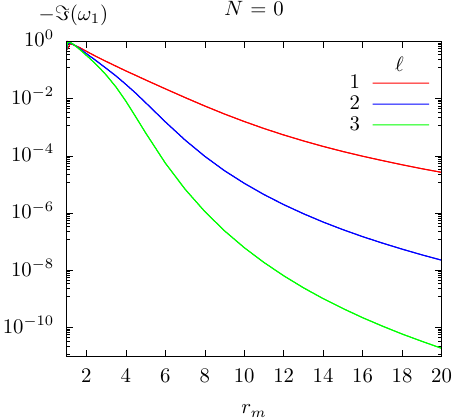}\hspace{12mm}\includegraphics[clip=true,width=0.326\textwidth]{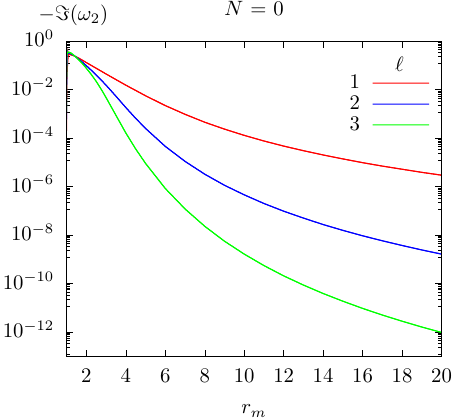}
\end{tabular}
\end{center}
\caption{\label{Fig_N0variousell} Real (top) and imaginary (bottom) parts of the Maxwell QNMs on Schwarzschild BHs in a cavity vs the mirror radius $r_m$, for $N=0$ and $\ell=1$ (red), $2$ (blue), $3$ (green), with the first (left) and second (right) boundary conditions. Note that in the bottom panels for imaginary part of QNMs with both boundary conditions, we use the semilogarithmic coordinates.}
\end{figure*}
\begin{figure*}
\begin{center}
\begin{tabular}{c}
\hspace{-4mm}\includegraphics[clip=true,width=0.326\textwidth]{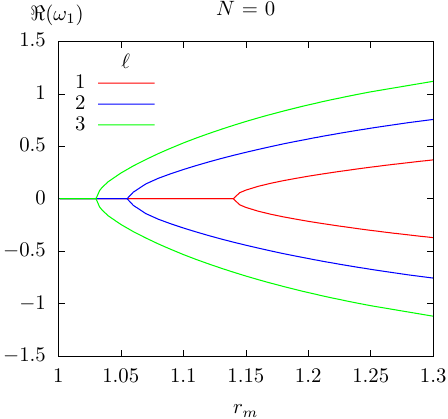}\hspace{12mm}\includegraphics[clip=true,width=0.326\textwidth]{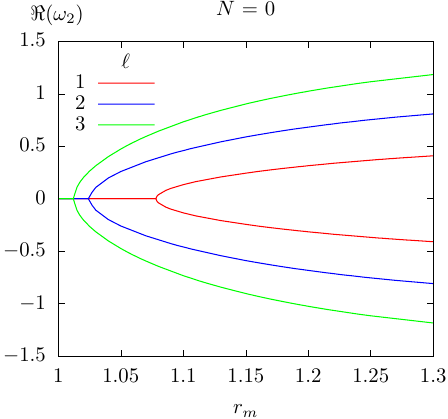}
\\
\vspace{2mm}
\\
\hspace{-2mm}\includegraphics[clip=true,width=0.316\textwidth]{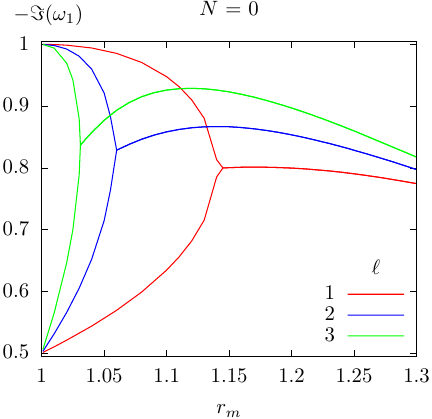}\hspace{14.5mm}\includegraphics[clip=true,width=0.316\textwidth]{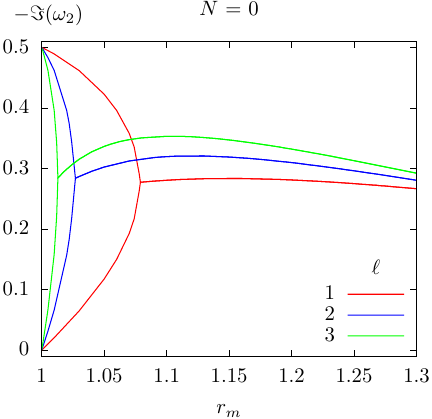}
\end{tabular}
\end{center}
\caption{\label{Fig_N0variousell_zoom} Spectrum bifurcation of the Maxwell fields on Schwarzschild BHs in a cavity for both boundary conditions with fixed $N$ but with different $\ell$. Note that this figure is obtained by zooming Fig.~\ref{Fig_N0variousell} around the event horizon. The critical mirror radius with the first (second) boundary condition are $r_{m1}^c \approx 1.14028$ ($r_{m2}^c \approx 1.07825$) for $\ell=1$, $r_{m1}^c \approx 1.05741$ ($r_{m2}^c \approx 1.02553$) for $\ell=2$ and $r_{m1}^c \approx 1.03079$ ($r_{m2}^c \approx 1.01270$) for $\ell=3$.}
\end{figure*}

We also explore the Maxwell spectrum in terms of $r_m$ for both boundary conditions with fixed $N$ but various $\ell$ in Fig.~\ref{Fig_N0variousell}, by taking $N=0$ and $\ell=1, 2, 3$ as examples. The general behavior is quite similar to Fig.~\ref{Fig_ell1variousN}. Again by zooming Fig.~\ref{Fig_N0variousell} around the event horizon, we have Fig.~\ref{Fig_N0variousell_zoom} which shows that the mode split effect does \textit{exist} not only for various values of $N$ but also for different values of $\ell$. By increasing $\ell$, shown in Figs.~\ref{Fig_N0variousell} and~\ref{Fig_N0variousell_zoom} for both boundary conditions, the magnitude of real part of QNMs always increases; while the magnitude of imaginary part decreases (increases) when $r_m$ is away from (around) $r_m^c$. Moreover, as shown in Fig.~\ref{Fig_N0variousell_zoom} and contrary to $N$ effects, the critical mirror radius for both boundary conditions decreases as $\ell$ increases, and by fixing all other parameters the critical mirror radius with the first boundary condition is again greater than the counterpart with the second boundary condition. In particular, the modes for different $\ell$ with both boundary conditions are \textit{degenerate}, as the mirror approaches the event horizon, which is again contradictory to $N$ effects.

\section{Discussion and Final Remarks}
\label{discussion}
In this paper, we have studied the Maxwell quasinormal spectrum on Schwarzschild BHs in a mirror-like cavity, by imposing the vanishing energy flux boundary conditions. We have shown that the vanishing energy flux principle is applicable to a BH-mirror system and leads to two distinct branches of modes. In particular, we demonstrated that the spectrum may \textit{bifurcate}, which shows, as an example among others, the similarity between BH-mirror and BH-AdS systems. More specifically, we unveiled that for both boundary conditions, when the mirror radius $r_m$ is less than the corresponding critical mirror radius $r_m^c$, the real part of the Maxwell QNMs \textit{vanishes} while the imaginary part \textit{branches off}. This feature is prominent when the mirror is placed close to the event horizon. We also found that, with fixed other parameters, the critical mirror radius with the first boundary condition is greater than the counterpart with the second boundary condition.

When the mirror radius is greater than and away from the corresponding critical mirror radius, we observed the following trends of the Maxwell spectrum. With fixed $\ell$ and $N$, the magnitude of real (imaginary) part of QNMs for both boundary conditions first increases and then decreases (increases) as the mirror radius $r_m$ approaches the event horizon $r_+$. With fixed $r_m$ and $\ell$, the magnitude of real (imaginary) part of QNMs for both boundary conditions increases (increases) as $N$ increases; while with fixed $r_m$ and $N$, the magnitude of real (imaginary) part of QNMs for both boundary conditions increases (decreases) as $\ell$ increases. When the mirror radius is around the critical mirror radius, on the other hand, the impact of $N$ on the real (imaginary) part of the spectrum is opposite (same) to the former case while the impact of $\ell$ on the real (imaginary) part of the spectrum is same (opposite) to the former case. Moreover, we observed that the critical mirror radius, for both boundary conditions, increases as $N$ increases but decreases as $\ell$ increases.

We also performed \textit{analytic} calculations of the Maxwell spectrum in some limits. In a pure cavity, the Maxwell normal modes were solved and we obtained two different sets of normal modes, which is contradictory to the AdS case where two sets of normal modes are the same up to one mode. For the case of BHs in a cavity, the Maxwell spectrum may be obtained in the low frequency approximations. By using a standard matching method, we got two distinct analytic expressions for the Maxwell QNMs, which matched well with numeric data.

By adding rotation to the background, superradiance and the corresponding instabilities may be triggered. It is then interesting to investigate superradiant instabilities in the BH-mirror system, and in particular to explore the interplay between superradiant instabilities and the spectrum bifurcation. Work along these directions is underway and we hope to report on them soon.

\bigskip

\noindent{\bf{\em Acknowledgements.}}
This work is supported by the National Natural Science Foundation of China under Grant Nos. 11705054, 11881240252, 12035005, and by the Hunan Provincial Natural Science Foundation of China under Grant No. 2018JJ3326.

\appendix
\section{The Maxwell normal modes in a cavity}
\label{app1}
In this appendix, we look for the Maxwell normal modes in the Regge-Wheeler-Zerilli formalism, by solving Eq.~\eqref{RWZeq} in a cavity \textit{analytically}, with vanishing energy flux boundary conditions given by Eqs.~\eqref{RWZbc2-1} and~\eqref{RWZbc2-2}.

In a pure cavity, the radial part of the Maxwell equations~\eqref{RWZeq} becomes
\begin{equation}
\left(\frac{d^2}{dr^2}+\omega^2-\dfrac{\ell(\ell+1)}{r^2}\right)\Psi(r)=0\;,\label{RWZeqCavity}
\end{equation}
which leads to the solution
\begin{equation}
\Psi(r) \sim \sqrt{r}\left(c_1\,J_{\ell+\frac{1}{2}}(\omega r)+c_2\,Y_{\ell+\frac{1}{2}}(\omega r)\right)\;, \label{CavitySol}
\end{equation}
where $J_\nu(z)$ and $Y_\nu(z)$ are the first kind and second kind Bessel functions, respectively.

By expanding the solution~\eqref{CavitySol} at the origin and by imposing regularity condition, we shall set $c_2=0$ and get 
\begin{equation}
\Psi(r) \sim \sqrt{r}J_{\ell+\frac{1}{2}}(\omega r)\;.\label{CavitySol2}
\end{equation}
Then the boundary condition~\eqref{RWZbc2-1} leads to 
\begin{equation}
\omega_1 r_m = j_{\ell+\frac{1}{2}, N}\;,\label{normalmodes1}
\end{equation}
while the boundary condition~\eqref{RWZbc2-2} leads to
\begin{equation}
\omega_2 r_m = \tilde{j}_{\ell+\frac{1}{2}, N}\;.\label{normalmodes2}
\end{equation}
Here $j_{\nu, N}$ and $\tilde{j}_{\nu, N}$ denote zeros of the Bessel function $J_\nu(\omega r)$ and zeros of $\partial_r[\sqrt{r}J_\nu(\omega r)]|_{r_m}$, and $N$ is the overtone number. 

As one may observe that two normal modes, obtained in Eqs~\eqref{normalmodes1} and~\eqref{normalmodes2}, are \textit{different}, which is contradictory to the AdS case where two normal modes are the same up to one mode~\cite{Wang:2015goa}. Moreover, these two normal modes obtained by imposing the vanishing energy flux boundary conditions are exactly the same with those obtained by imposing the conductor condition~\cite{Brito:2015oca}.

By adding a BH in a cavity and when the BH size is much smaller than the mirror radius, the Maxwell QNMs may be solved perturbatively on top of normal modes obtained in a pure cavity. Such calculation may be only achieved in the Teukolsky formalism. In the next appendixes, we briefly present the Maxwell equations in the Teukolsky formalism and illustrate the corresponding explicit boundary conditions.     

\section{The Maxwell equations in the Teukolsky formalism}
\label{app2}
By using the Newmann-Penrose algorithm~\cite{Newman:1961qr}, the radial part of the Maxwell equations may be obtained alternatively in the Teukolsky formalism~\cite{Teukolsky:1973ha}
\begin{equation}
\Delta_r^{-s}\dfrac{d}{dr}\left(\Delta_r^{s+1}\dfrac{d R_{s}(r)}{dr}\right)+H(r)R_{s}(r)=0\;,\label{Teukeq}
\end{equation}
with the spin parameter $s=\pm1$, and where
\begin{eqnarray}
H(r)=\dfrac{K_r^2-i s K_r \Delta_r^\prime}{\Delta_r}+2is K_r^\prime +\dfrac{s+|s|}{2}\Delta_r^{\prime\prime}
-\lambda\;.\nonumber
\end{eqnarray}
Note that here we have 
\begin{equation}
\Delta_r=r^2f(r)\;,\;\;\;\quad K_r=\omega r^2\;,\;\;\;\quad\lambda=\ell(\ell+1)\;,\nonumber
\end{equation}
where the metric function $f(r)$ is given in Eq.~\eqref{metricfunc}.

\section{Boundary conditions of the Teukolsky variables}
\label{app3}
To solve the Maxwell QNMs in the Teukolsky formalism, one has to solve Eq.~\eqref{Teukeq} with the corresponding boundary conditions. To be specific, here we apply the generic vanishing energy flux principle to the Teukolsky variables with spin $s=-1$, and then one obtains~\cite{Wang:2015goa,Brito:2015oca}
\begin{equation}
R^\prime_{-1}(r_m)=\left(\dfrac{1}{r_m}+\dfrac{i\omega}{f(r_m)}\right)R_{-1}(r_m)\;,\label{Teukolbc1}
\end{equation}
and
\begin{equation}
R^\prime_{-1}(r_m)=\left(\dfrac{1}{r_m}-\dfrac{i\lambda}{\omega r_m^2}+\dfrac{i\omega}{f(r_m)}\right)R_{-1}(r_m)\;,\label{Teukolbc2}
\end{equation}
where $\prime$ denotes a derivative with respect to $r$, the location of a mirror is represented by $r_m$, the metric function $f(r)$ is given by Eq.~\eqref{metricfunc}, and
\begin{equation}
R^\prime_{-1}(r_m)\equiv\dfrac{dR_{-1}(r)}{dr}|_{r=r_m}\;,\;\;\;\;\;\;\lambda=\ell(\ell+1)\;.\nonumber
\end{equation}

\section{Analytic calculations}
\label{secana}
In the low frequency limit, the Maxwell quasinormal spectrum for Schwarzschild BHs in a cavity may be calculated \textit{analytically} when the mirror is placed far away from the hole. To this end, one shall solve the Maxwell equations in the Teukolsky formalism, which is given by Eq.~\eqref{Teukeq} presented in the appendix~\ref{app2}, with the corresponding boundary conditions given by Eqs.~\eqref{Teukolbc1} and~\eqref{Teukolbc2} outlined in the appendix~\ref{app3}. 

Such analytic calculations can be preformed following the standard matching method. For this purpose, we shall first divide the spacetimes outside the Schwarzschild BH into \textit{near} and \textit{far} regions, which are defined as $r-r_+\ll1/\omega$ and $r\gg r_+$, respectively. In the low frequency approximation $\omega r_+\ll1$, the near region and the far region are overlapped, and the Maxwell spectrum my be obtained perturbatively on top of the normal modes for a pure cavity. 

\subsection{Near region solution}
To be specific but without loss of generality, we take the spin parameter $s=-1$ in the Teukolsky equation given by Eq.~\eqref{Teukeq}. Then the Teukolsky equation of the Maxwell field in the near region, i.e. $r-r_+\ll1/\omega$, becomes
\begin{align}
\Delta_rR_{-1}''+\left[\dfrac{\omega r_+^4+i\omega r_+^2(2r-r_+)}{\Delta_r}-\lambda\right]R_{-1}=0\;,\label{nearEq}\end{align}
where
\begin{equation}
R_{-1}\equiv R_{-1}(r)\;,\;\;\;\Delta_r=r^2f(r)\;,\;\;\;\lambda=\ell(\ell+1)\;,
\end{equation}
and where $f(r)$ is the metric function given in Eq.~\eqref{metricfunc}.

Now by defining a new dimensionless variable $z$
\begin{equation}
z=1-\dfrac{r_+}{r}\;,\nonumber
\end{equation}
and by introducing a new function $F(z)$
\begin{equation}
R_{-1}(z)=z^{i\omega r_+}(1-z)^{\ell}F(z)\;,\label{transformRF}
\end{equation}
Eq.~\eqref{nearEq} may be transformed into a standard hypergeometric equation
\begin{align}
z(1-z)F''(z)+[c-(1+a+b)z]F'(z)-abF(z)=0\;,\label{nearEq2}
\end{align}
where
\begin{equation}
a=\ell\;,\;\;\;b=\ell+1+2i\omega r_+\;,\;\;\;c=2i\omega r_+\;.
\end{equation}
By imposing an ingoing wave boundary condition at the horizon, from Eqs.~\eqref{nearEq2} and~\eqref{transformRF}, one obtains 
\begin{equation}
R_{-1}(z)\sim z^{1-i\omega r_+}{(1-z)^{\ell}}\;{_2F_1}(1+a-c,1+b-c,2-c,z)\;,\label{nearsol}
\end{equation}
where $_2F_1(a, b, c, z)$ is the standard hypergeometric function.

To employ the matching method, we shall first expand the above near region solution at large $r$. For this purpose, by taking $z\to1$ and by using the property of the hypergeometric function, we get
\begin{equation}
R_{-1}(r)\sim \Gamma(2-2i\omega r_+)\left[\dfrac{\mathcal{A}_1}{r^{\ell}}+\mathcal{A}_2r^{\ell+1}\right]\;,\label{nearsolfar}
\end{equation}
where
\begin{align}
&\mathcal{A}_1=\dfrac{\Gamma(-2\ell-1)r_+^{\ell}}{\Gamma(1-\ell)\Gamma(-\ell-2i\omega r_+)}\;,\nonumber
\\
&\mathcal{A}_2=\dfrac{\Gamma(2\ell+1)r_+^{-\ell-1}}{\Gamma(\ell+2)\Gamma(\ell+1-2i\omega r_+)}\;.\label{nearexp}
\end{align}
\subsection{Far region solution}
In the far region $r\gg r_+$, the BH effects may be neglected ($M\to0$) so that the Teukolsky equation~\eqref{Teukeq} with $s=-1$ turns into
\begin{equation}
r^2R_{-1}''(r)-\Big[\omega r(2i-\omega r)+\ell(\ell+1)\Big]R_{-1}(r)=0\;.\label{fareq}
\end{equation}
The general solution of Eq.~\eqref{fareq} is given by 
\begin{equation}
R_{-1}(r)\sim\alpha M_{-1,\ell+\frac{1}{2}}(2i\omega r)+\beta W_{-1,\ell+\frac{1}{2}}(2i\omega r)\;,\label{farsol}
\end{equation}
where $M_{\kappa, \mu}(z)$ and $W_{\kappa, \mu}(z)$ are Whittaker functions, and the two integration constants $\alpha, \beta$ should be determined by boundary conditions.

By imposing the first boundary condition given by Eq.~\eqref{Teukolbc1}, one obtains
\begin{equation}
\alpha\ell M_{0,\ell+1/2}(2i\omega r_m)-\beta W_{0,\ell+1/2}(2i\omega r_m)=0\;,\nonumber
\end{equation}
which implies 
\begin{equation}
\dfrac{\beta}{\alpha}=\dfrac{lM_{0,\ell+1/2}(2i\omega r_m)}{W_{0,\ell+1/2}(2i\omega r_m)}\;.\label{alphabetabc1}
\end{equation}
While by imposing the second boundary condition given by Eq.~\eqref{Teukolbc2}, one gets
\begin{align}
&\alpha[i\ell(\ell+1)M_{-1,\ell+1/2}(2i\omega r_m)+\ell\omega r_mM_{0,\ell+1/2}(2i\omega r_m)]\nonumber\\
&+\beta[i\ell(\ell+1)W_{-1,\ell+1/2}(2i\omega r_m)-\omega r_mW_{0,\ell+1/2}(2i\omega r_m)]\nonumber\\
&=0\;,\nonumber
\end{align}
which indicates
\begin{equation}
\dfrac{\beta}{\alpha}=\dfrac{\ell\mathcal{C}_1-\ell(\ell+1)\mathcal{C}_2}{\tilde{\mathcal{C}}_1+\ell(\ell+1)\tilde{\mathcal{C}}_2}\;,\label{alphabetabc2}
\end{equation}
where
\begin{align}
&\mathcal{C}_1=i\omega r_mM_{0,\ell+1/2}(2i\omega r_m)\;,\nonumber
\\
&\tilde{\mathcal{C}}_1=i\omega r_mW_{0,\ell+1/2}(2i\omega r_m)\;,\nonumber
\\
&\mathcal{C}_2=M_{-1,\ell+1/2}(2i\omega r_m)\;,\nonumber
\\
&\tilde{\mathcal{C}}_2=W_{-1,\ell+1/2}(2i\omega r_m)\;.\label{Cexp}
\end{align}

To match the near region solution, by expanding the far region solution at small $r$, we obtain
\begin{equation}
R_{-1}(r)\sim\dfrac{\mathcal{B}_1}{r^{\ell}}+\mathcal{B}_2r^{\ell+1}\;,\label{farsolnear}
\end{equation}
where
\begin{align}
&\mathcal{B}_1=\beta\dfrac{\Gamma(2\ell+1)}{\Gamma(\ell+2)}(2i\omega)^{-\ell}\;,\nonumber
\\
&\mathcal{B}_2=\alpha(2i\omega)^{\ell+1}\;.\label{farexp}
\end{align}

\subsection{Overlap region}
In the low frequency approximations, the near and far regions are overlapped so that the solutions obtained in these two regions may be matched in the intermediate region. From Eqs.~\eqref{nearsolfar} and~\eqref{farsolnear}, and by using the matching condition $\mathcal{A}_1\mathcal{B}_2=\mathcal{A}_2\mathcal{B}_1$, one obtains
\begin{equation}
\dfrac{\beta}{\alpha}=-(i\sigma)^{2\ell+2}\dfrac{(\ell-1)!}{2(2\ell+1)!}\left(\dfrac{(\ell+1)!}{(2\ell)!}\right)^2\prod_{k=1}^{\ell}(k^2+\sigma^2)\;,\label{matchingeq}
\end{equation}
where $\sigma=2\omega r_+$.

Given the relations between $\alpha$ and $\beta$ in Eqs.~\eqref{alphabetabc1} and~\eqref{alphabetabc2} obtained from boundary conditions and the matching condition~\eqref{matchingeq}, for the first boundary we have
\begin{align}
&-(i\sigma)^{2\ell+2}\dfrac{(\ell-1)!}{2(2\ell+1)!}\left(\dfrac{(\ell+1)!}{(2\ell)!}\right)^2\prod_{k=1}^{\ell}(k^2+\sigma^2)\nonumber
\\
&=\dfrac{lM_{0,\ell+1/2}(2i\omega r_m)}{W_{0,\ell+1/2}(2i\omega r_m)}\;,\label{matchingbc1}
\end{align}
while for the second boundary we have 
\begin{align}
&-(i\sigma)^{2\ell+2}\dfrac{(\ell-1)!}{2(2\ell+1)!}\left(\dfrac{(\ell+1)!}{(2\ell)!}\right)^2\prod_{k=1}^{\ell}(k^2+\sigma^2)\nonumber
\\
&=\dfrac{\ell\mathcal{C}_1-\ell(\ell+1)\mathcal{C}_2}{\tilde{\mathcal{C}}_1+\ell(\ell+1)\tilde{\mathcal{C}}_2}\;.\label{matchingbc2}
\end{align}

In the low frequency approximations, the left hand side of Eqs.~\eqref{matchingbc1} and~\eqref{matchingbc2} are $\mathcal{O}((\omega r_+)^{2\ell+2})$. Therefore, we define $\omega_{1, N}$ and $\omega_{2, N}$ as zeros of the right hand side of Eqs.~\eqref{matchingbc1} and~\eqref{matchingbc2} and here $N$ is the overtone number. These two modes are normal modes for a pure cavity with the first and second boundary, and they are the same with the corresponding modes obtained in the Regge-Wheeler-Zerilli formalism, given by Eqs.~\eqref{normalmodes1} and~\eqref{normalmodes2}. 

To get a complete solution, we further assume 
\begin{equation}
\omega_j=\omega_{j, N}+i\delta_j\;,\label{omegaexp}
\end{equation}
where $j= 1, 2$, corresponding to the first and second boundary, and $\delta_j$ is a small quantity by comparing with $\omega_{j, N}$. 

By substituting Eq.~\eqref{omegaexp} into Eq.~\eqref{matchingbc1}, and considering $M_{0,\ell+1/2}(2i\omega_{1, N} r_m)=0$, one obtains $\delta_1$
\begin{equation}
\delta_1\simeq\dfrac{\omega_{1,N}r_mW_{0,\ell+1/2}(2i\omega_{1,N}r_m)}{i\ell(\ell+1)M_{1,\ell+1/2}(2i\omega_{1,N}r_m)}\gamma_1\;,
\end{equation} 
where
\begin{align}
\gamma_1=&-(2i\omega_{1, N}r_+)^{2\ell+2}\dfrac{(\ell-1)!}{2(2\ell+1)!}\left(\dfrac{(\ell+1)!}{(2\ell)!}\right)^2\times\nonumber
\\
&\times\prod_{k=1}^{\ell}(k^2+4\omega_{1, N}^2r_+^2)\;.\nonumber
\end{align}

For the second boundary, by substituting Eq.~\eqref{omegaexp} into Eq.~\eqref{matchingbc2} and considering 
\begin{align}
&i\omega_{2, N} r_mM_{0,\ell+1/2}(2i\omega_{2, N} r_m)\nonumber\\=&(\ell+1)M_{-1,\ell+1/2}(2i\omega_{2, N} r_m)\;,\nonumber
\end{align}
$\delta_2$ may be obtained 
\begin{equation}
\delta_2\simeq\dfrac{\omega_{2,N}r_m\left(\ell(\ell+1)\tilde{\mathcal{C}}_2(\omega_{2,N})
+\tilde{\mathcal{C}}_1(\omega_{2,N})\right)}{i\ell(\ell+1)\left(\mathcal{C}_3-\ell\mathcal{C}_4\right)}\gamma_2\;,
\end{equation}
where $\tilde{\mathcal{C}}_1(\omega_{2,N})$ and $\tilde{\mathcal{C}}_2(\omega_{2,N})$ are given in Eq.~\eqref{Cexp}, and
\begin{align}
\mathcal{C}_3=&i\omega_{2,N}r_mM_{1,\ell+1/2}(2i\omega_{2,N}r_m)\;,\nonumber
\\
\mathcal{C}_4=&M_{0,\ell+1/2}(2i\omega_{2,N}r_m)\;,\nonumber
\\
\gamma_2=&-(2i\omega_{2, N} r_+)^{2\ell+2}\dfrac{(\ell-1)!}{2(2\ell+1)!}\left(\dfrac{(\ell+1)!}{(2\ell)!}\right)^2\times\nonumber\\&\times\prod_{k=1}^{\ell}(k^2+4\omega_{2, N}^2r_+^2)\;.
\end{align}
Note that $\delta_1$ and $\delta_2$, in general, are complex and the real part of them, $i.e.$ $\Re(\delta_1)$ and $\Re(\delta_2)$, reflects the damping of the QNMs.

\bibliographystyle{h-physrev4}
\bibliography{BHMirror}
\end{document}